\documentclass[a4paper]{jpconf}
\usepackage{geometry}
\usepackage{amsfonts}
\usepackage{amsmath}
\usepackage{graphicx}
\usepackage{dcolumn}
\usepackage{bm}
\usepackage{slashed}
\usepackage{amssymb}
\usepackage{MnSymbol}
\usepackage{xcolor}
\usepackage{verbatim}
\usepackage{todonotes}
\usepackage{grffile}
\usepackage{cite}
\usepackage{subfigure}
\usepackage{hyperref}

\usepackage{soul}

\begin{document}
\title{Quark mass effects in quark number susceptibilities}

\author{Thorben Graf$^1$, Peter Petreczky$^2$}
\address{$^1$ SUBATECH, UMR 6457, Universit́e de Nantes, Ecole des Mines de Nantes, IN2P3/CNRS. 4 rue Alfred Kastler, 44307 Nantes cedex 3, Franc}
\address{$^2$ Physics Department, Brookhaven National Laboratory, Upton, NY 11973, USA}

\ead{thorben\underline{ }graf@gmx.de}

\begin{abstract}
  The quark degrees of freedom of the QGP with special focus on mass effects are investigated. A next-to-leading-order perturbation theory approach with quark mass dependence is applied and compared to lattice QCD results.
\end{abstract}

%%%%%%%%%%%%%%%%%%%%%%%%%%%%%%%%%%%%%%%%%%%%%%%%%%%%%%%%%%%%%%%
\section{Introduction}

The experiments at the Relativistic Heavy Ion Collider (RHIC) at Brookhaven National Laboratory as well as at the Large Hadron Collider (LHC) in Geneva showed clear evidences for the creation of a novel state of matter that is usually refered to as Quark-Gluon-Plasma (QGP) (see e.g. Ref.~\cite{Bala:2016hlf} for a recent review). According to asymptotic freedom, at sufficiently high $T$ the quarks and gluons in this medium are weakly coupled and enable the description of the underlying physics by perturbation theory. On the other hand Lattice Monte Carlo simulations are the method of choice to fully capture the thermodynamic effects of strongly interacting matter at least for nonvanishing chemical potential. In this work these two approaches are used to investigate important thermodynamical quantities such as quark number fluctuations that also can be accessed by experiments. The focus lies on the incorporation of quark masses in the calculations.

%%%%%%%%%%%%%%%%%%%%%%%%%%%%%%%%%%%%%%%%%%%%%%%%%%%%%%%%%%%%%%%
\section{Calculation}

For the perturbative calculations we resort to a next-to-leading order setup including mass dependence that was intensively described in Ref.~\cite{Graf:2015tda}. The lattice data to which the perturbative results are compared to can be found in Refs.~\cite{Petreczky:2009cr,Bazavov:2013uja,Ding:2015fca}. \\
A few words about the chosen accuracy in perturbation theory are in order. The convergence of the weak coupling expansion is very bad. This behaviour is already known for decades. The free energy of hot QCD can in fact be separated into contributions from momentum scales $T$, $gT$ and $g^2T$ \cite{Braaten:1995jr}, that are called hard, soft and supersoft modes. This is why the pressure of hot QCD is usually written as $p_{\text{QCD}}\equiv p_{\text{hard}}+p_{\text{soft}}$. It can be shown that $p_{\text{hard}}$ is the contribution from the scale $\sim T$ and has a well behaved perturbative expansion. This is not the case for $p_{\text{soft}}$ where the contributions from the soft and supersoft scales $gT$, $g^2T$ are included. However, in this work the impact of the quark masses is studied. The convergence of the expansion in terms of the mass dependece is different, which was nicely illustrated in Ref.~\cite{Laine:2006cp}. This is why the accuracy of the perturbative order is restricted to $g^2$ here. Accidentaly, the $g^2$ perturbative result is close to the full non-perturbative result and therefore it serves as a reasonable starting point to study quark mass effects.

%%%%%%%%%%%%%%%%%%%%%%%%%%%%%%%%%%%%%%%%%%%%%%%%%%%%%%%%%%%%%%%
\section{Results}

\begin{center}
  \begin{figure}
    \subfigure[]{
     \includegraphics[width=8cm]{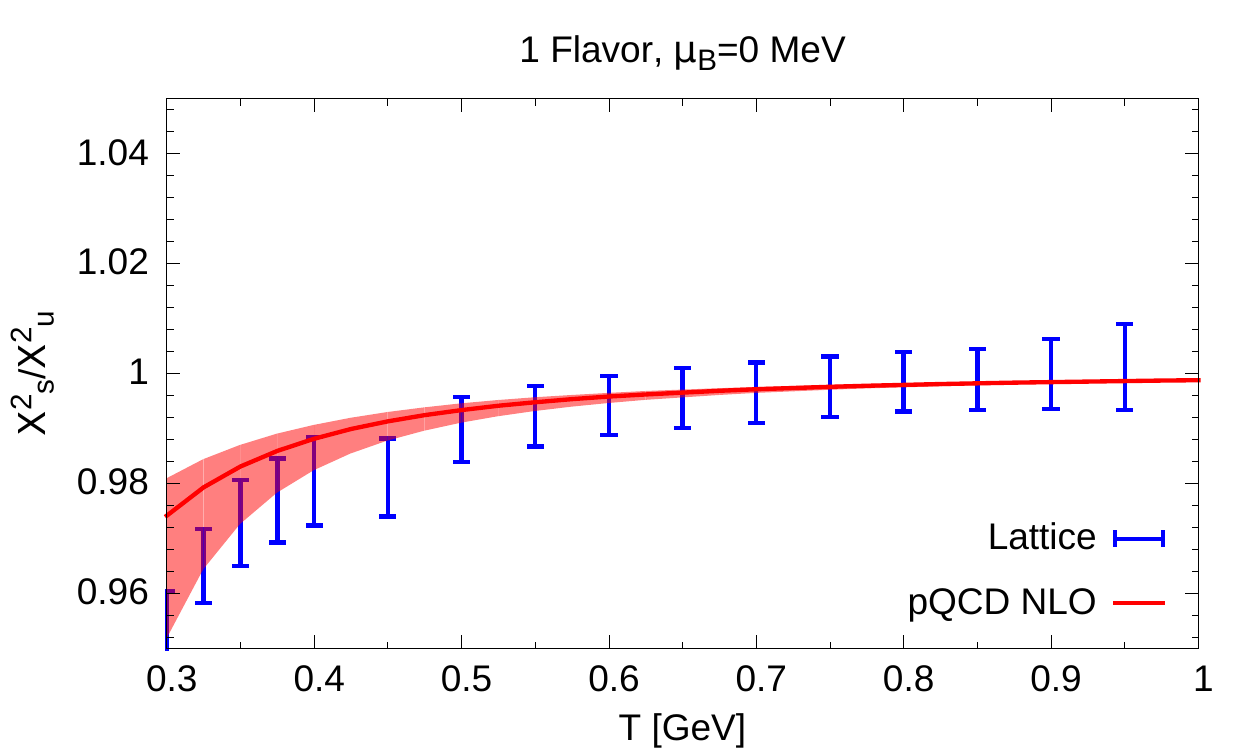}}
    \subfigure[]{
      \includegraphics[width=8cm]{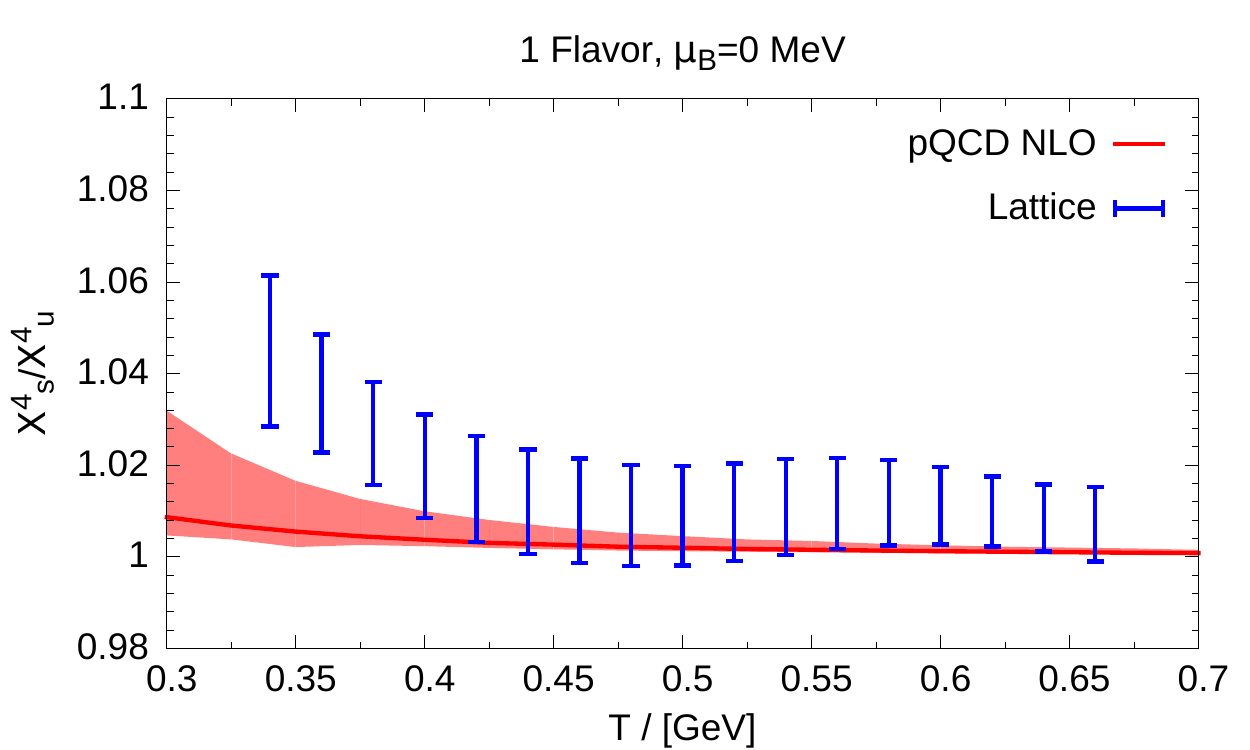}}
   %\caption{Comparison of the ratios $\chi_s^2/\chi_u^2$ (a) and $\chi_s^4/\chi_u^4$ (b) at $\mu_B=0$ with lattice results, denoted by Lattice, \cite{Petreczky:2009cr}.}
    \subfigure[]{
     \includegraphics[width=8cm]{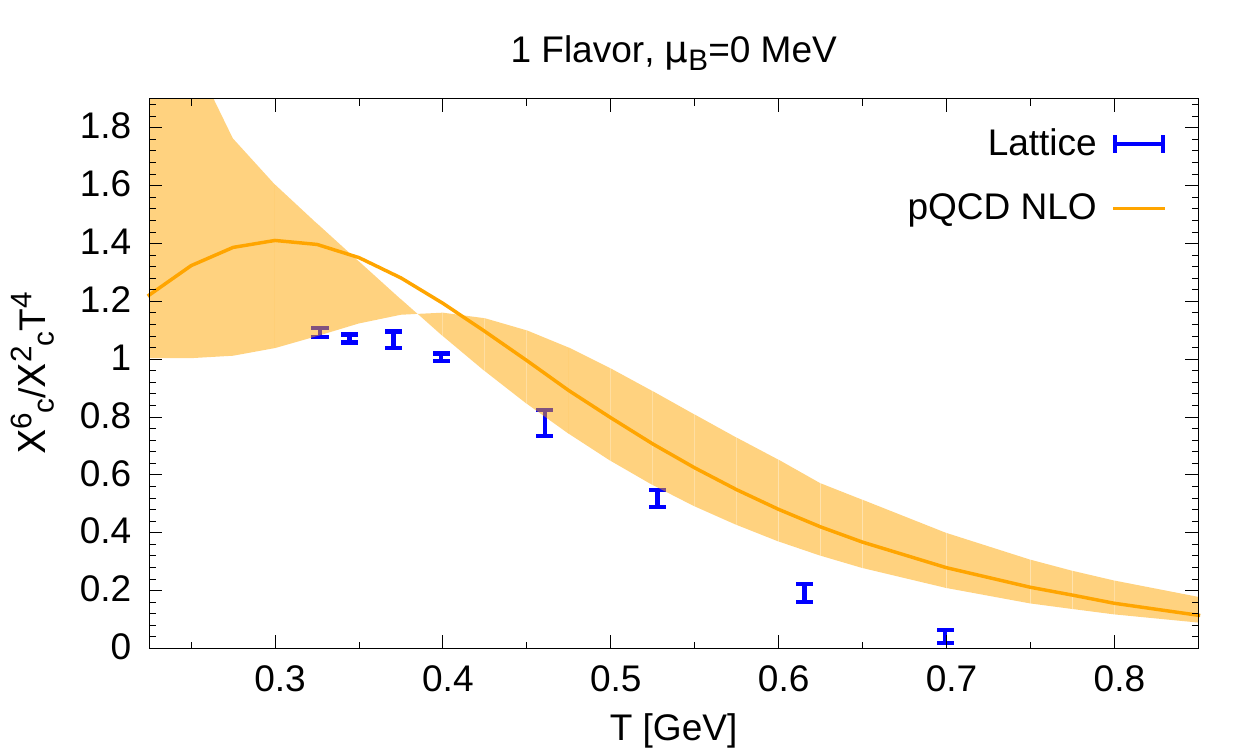}}
    \subfigure[]{
      \includegraphics[width=8cm]{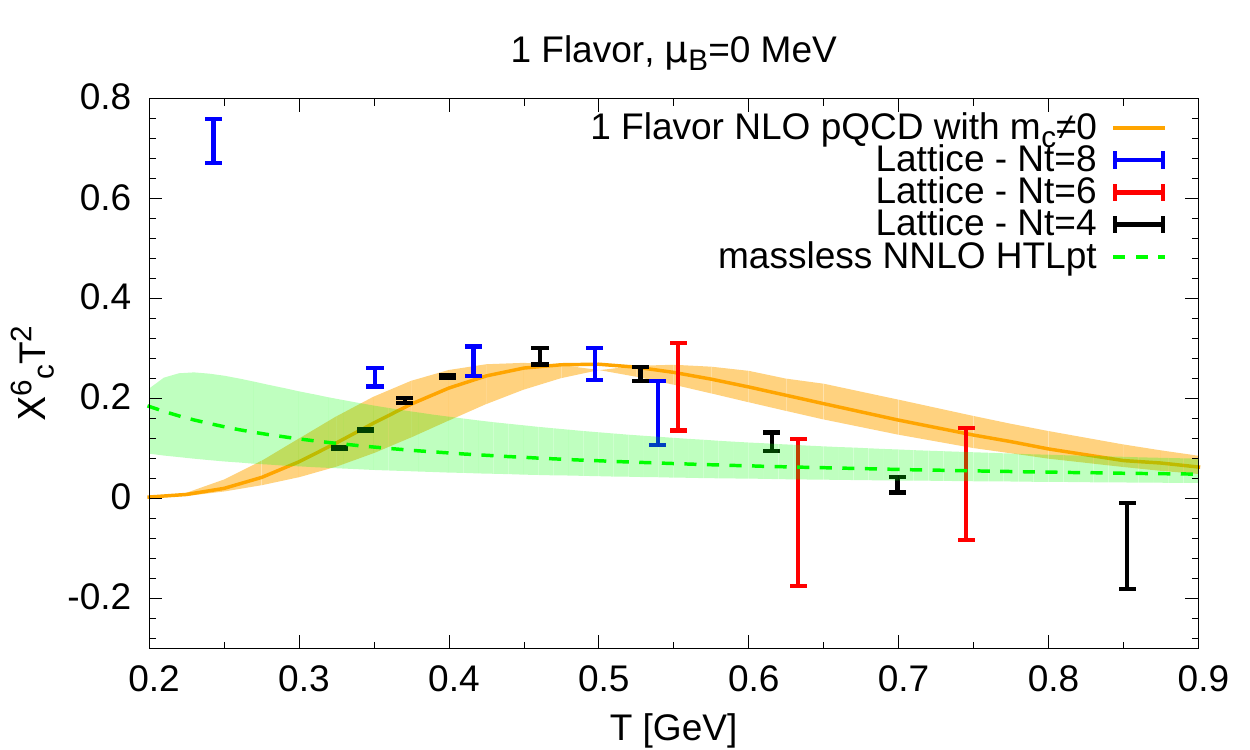}}
   \caption{Comparison of the ratios $\chi_s^2/\chi_u^2$ (a), $\chi_s^4/\chi_u^4$ (b) and $\chi_c^6/\chi_c^2\cdot T^4$ (c) at $\mu_B=0$ with lattice QCD results \cite{Petreczky:2009cr,Bazavov:2013uja,Ding:2015fca}. The comparison of $\chi_c^6\cdot T^2$ at $\mu_B=0$ with lattice QCD results \cite{Petreczky:2009cr} and HTLpt results \cite{Haque:2014rua} is illustrated in (d).}
    \label{fig:Chis}
  \end{figure}
\end{center}

In all numerical perturbative calculations the light quark masses are set to $m_u=2.3 \text{ MeV}$ and $m_d=4.8 \text{ MeV}$. The setup of Ref.~\cite{Graf:2015tda} includes a running prescription of the coupling constant. Such a behaviour usually appears at NNLO so that higher order effects are partially taken into account though the calculation of the thermodynamic potential is at NLO. As in Ref.~\cite{Fraga:2004gz} we choose $\alpha_s(\Lambda)=\frac{4\pi}{\beta_0L}\left[1-2\frac{\beta_1}{\beta_0^2}\frac{\ln L}{L}\right]$, with $\Lambda$ being the renormalization scale and $L=2\ln(\Lambda/\Lambda_{\overline{\text{MS}}})$, $\beta_0=11-2N_f/3$ and $\beta_1=51-19N_f/3$. For the heavier quarks we take into account the running $m_s(\Lambda)=\hat{m}_s\left(\frac{\alpha_s}{\pi}\right)^{4/9}\left[1+0.895062\frac{\alpha_s}{\pi}\right]$ and $m_c(\Lambda)=\hat{m}_c\left(\frac{\alpha_s}{\pi}\right)^{12/25}\left[1+1.01413\frac{\alpha_s}{\pi}\right]$. The scale $\Lambda_{\overline{\text{MS}}}$ is fixed by requiring $\alpha_s\simeq0.336$ $\Lambda=1.5$ GeV \cite{Bazavov:2014soa}; one obtains $\Lambda_{\overline{\text{MS}}}\simeq343$ MeV for 2+1 flavors and $\Lambda_{\overline{\text{MS}}}\simeq285$ MeV for 2+1+1 flavors. Requiring $m_s=93.6$ at $\Lambda=2$ GeV \cite{Chakraborty:2014aca} for the strange quark mass, leads to $\hat{m}_s=252$ MeV. For calculations that include $m_c$, we proceeded in the following way. For the calculations we use either 2+1 flavor running or 2+1+1 flavor running depending on the temperature. At $\Lambda=m_c^{\text{pole}}\eqsim1.7$ GeV we change between the different number of flavors. Therefore the 2+1-running $\alpha_s$ is again fixed by $\alpha_s\simeq0.336$ $\Lambda=1.5$ GeV. The $\Lambda_{\overline{\text{MS}}}$ for the 2+1+1-running is determined by matching them at the scale of the c-quark pole mass, which means, that $\alpha_s^{2+1}(\Lambda=m_c^{\text{pole}})=\alpha_s^{2+1+1}(\Lambda=m_c^{\text{pole}})$. It yields $\Lambda^{2+1}_{\overline{\text{MS}}}\simeq343$ MeV and $\Lambda^{2+1+1}_{\overline{\text{MS}}}\simeq285$ MeV. Of course for $m_c$ the analogous condition must be fulfilled, $m_c^{2+1}(\Lambda=m_c^{\text{pole}})=m_c^{2+1+1}(\Lambda=m_c^{\text{pole}})$, which means for $\hat{m}_c^{2+1}=3231$ MeV and $\hat{m}_c^{2+1+1}=3507$ MeV. With these conventions, the only freedom left is the choice of $\Lambda$ which is chosen to be $\Lambda=2\pi T$. For this choice of the renormalization scale the matching point of 2+1 and 2+1+1 flavor running corresponds to roughly $T=271$ MeV. The band uncertainties in our plots shown below result from the variation of $\Lambda$ by a factor of two. \\
We first calculated the pressure and investigated the impact of 2+1+1 compared to 2+1 flavors. The lattice data lay completely in the uncertainty band of the perturbative calculation. However, at low temperatures this uncertainty was quite large, so that one can conclude that for temperatures $T\gtrsim 300$ MeV both methods are in good agreement. The contribution of the charm quarks to the pressure revealed the Boltzmann-suppression for low temperatures, as expected. In this approximation the distribution functions are simply replaced by the Boltzmann distribution, and at leading order the partical charm pressure is $p_c(T,\mu_c)=f(T)\cdot\cosh(\mu_c/T)$, where $\mu_c$ is the quark chemical potential of the charm-quark. \\
In Ref.~\cite{Graf:2015tda} it was pointed out that the susceptibilities are the thermodynamic quantities that are the most sensitive ones to the quark mass effects. This is why we are especially interested in investigations of these quantities. We compute quark number susceptibilities, which are defined as derivatives with respect to the corresponding quark chemical potentials $\vec{\mu}\equiv(\mu_u,\mu_d,\dots,\mu_{N_f})$ as
\begin{equation}  
  \left.\chi_{ijk\dots}(T)\equiv\frac{\partial^{i+j+k+\dots}p(T,\vec{\mu})}{\partial\mu^i_u\partial\mu^j_d\partial\mu^k_s\dots}\right|_{\vec{\mu}=0}.
\end{equation}
From the perturbative point of view the quark mass depedence of the quark number susceptibilities has not been investigated. We start with the ratio of the strange- over up-quark susceptibility of second order that is depicted in Fig.~\ref{fig:Chis} (a) and compared with lattice data of Ref.~\cite{Bazavov:2013uja}. The trend is obviously described by both of the approaches in the same way even at temperatures down to $T=350$ MeV. Only at even lower temperatures there is a slight deviation that is catched by the uncertainties though. Note that previous perturbative calculations would only result in a horizontal line at 1 because there is no difference between $\chi_u^2$ and $\chi_s^2$. The same ratio but this time at fourth order is shown in Fig.~\ref{fig:Chis} (b). Again the trend of the lattice data \cite{Ding:2015fca} is in good agreement with the perturbative results. For low temperatures, $T<400$ MeV, there is an increasing deviation. The tiny maximun provided by the lattice data for temperatures of roughly $T=580$ MeV cannot be reproduced by the perturbative setup.

In Fig.~\ref{fig:Chis} (c) the ratio of the sixth- and second order charm-quark susceptibilities is studied. Having a look at the expression for $p_c$ we expect the ratio to be close to 1 at least at leading order. Obviously the perturbative result in Fig.~\ref{fig:Chis} (c) is slightly above 1. This behaviour can be explained if we have a more detailed look at the NLO expressions. At NLO products of distribution functions appear. If the Boltzmann-approximation is applied a term proportional to $\cosh(2\beta\mu_c)$ and a term independent of $\mu_c$ are arising. For the normalized ratio $\chi_c^6/\chi_c^2\cdot T^4$ this means it is proportional to $\frac{\cosh(\beta\mu_c)+2^6\cosh(2\beta\mu_c)}{\cosh(\beta\mu_c)+2^2\cosh(2\beta\mu_c)}$ which is obviously bigger than 1. In general the comparison of this ratio with lattice data is successful because the trend is described analogously even when the perturbative results are a little bit above the lattice data.

The sixth order charm-quark susceptibility is interesting by itself because lattice simulations showed that there is a maximum around $T=450$ MeV that cannot be reproduced by even resummed three-loop perturbative calculations without mass dependence \cite{Haque:2014rua}. This is why it was not clear so far if the bump is caused by nonvanishing bare quark masses or by higher order effects beyond three-loop. Our results (solid line) are shown in Fig.~\ref{fig:Chis} (d) and compared to lattice data at different temporal spacings and NNLO HTLpt (dashed line). Obviously our approach is able to recover the maximum that is also described by the lattice calculations. We did calculations at LO where the bump also appeared, but it lay outside of the lattice error bars. For the NLO result the bump is within the error bars of lattice data at $N_t=8$. It is also clearly visible that the Boltzmann-approximation holds for our calculation at low temperatures because of the large charm-quark mass of roughly $m_c=1.2$ GeV at a temperature of $T=200$ MeV.

%%%%%%%%%%%%%%%%%%%%%%%%%%%%%%%%%%%%%%%%%%%%%%%%%%%%%%%%%%%%%%%
\section{Summary}

The effects of heavy quark flavors on the pressure and susceptibilities have been investigated. A perturbative and a lattice Monte Carlo approach were compared in order to see if mass effects can be reproduced in the same way. The considered ratios of the susceptibilities show a good agreement between the two methods. The investigation included calculations of the quark number susceptibility up to the sixth order which confirmed the behaviour already seen in previous lattice calculations. An obvious following project would be the inclusion of higher orders in the perturbative setup.

%%%%%%%%%%%%%%%%%%%%%%%%%%%%%%%%%%%%%%%%%%%%%%%%%%%%%%%%%%%%%%%
\section*{Acknowledgments}

The authors want to thank Moritz Greif, Swagato Mukherjee, Robert D. Pisarski and Sayantan Sharma for fruitful discussions. T.G. was supported by the Helmholtz International Center for FAIR, the Helmholtz Graduate School HGS-HIRe, the  Service pour la science et la technologie pr\`{e}s l'Ambassade de France en Allemagne and Campus France. T.G. is grateful for the kind hospitality of the Nuclear Theory Group at Brookhaven National Laboratory, where part of this work was carried out. P.P. was supported by U.S. Department of Energy under contract No. DE-SC0012704 \\

%%%%%%%%%%%%%%%%%%%%%%%%%%%%%%%%%%%%%%%%%%%%%%%%%%%%%%%%%%%%%%%

%%%%%%%%%%%%%%%%%%%%%%%%%%%%%%%%%%%%%%%%%%%%%%%%%%%%%%%%%%%%%%%
\section*{References}

%\bibliographystyle{ieeetr}
%\bibliography{Charm_DoF_Bib}
%\bibliographystyle{plain}

\end{document}